\documentclass[12pt,epsf]{article}
\pdfoutput=1
\usepackage{graphicx}
\usepackage{amssymb}
\usepackage{amsmath}
\begin{document}

\def\a{\alpha}
\def\b{\beta}
\def\c{\varepsilon}
\def\d{\delta}
\def\e{\epsilon}
\def\f{\phi}
\def\g{\gamma}
\def\h{\theta}
\def\k{\kappa}
\def\l{\lambda}
\def\m{\mu}
\def\n{\nu}
\def\p{\psi}
\def\q{\partial}
\def\r{\rho}
\def\s{\sigma}
\def\t{\tau}
\def\u{\upsilon}
\def\v{\varphi}
\def\w{\omega}
\def\x{\xi}
\def\y{\eta}
\def\z{\zeta}
\def\D{\Delta}
\def\G{\Gamma}
\def\H{\Theta}
\def\L{\Lambda}
\def\F{\Phi}
\def\P{\Psi}
\def\S{\Sigma}

\def\o{\over}
\def\beq{\begin{eqnarray}}
\def\eeq{\end{eqnarray}}
\newcommand{\gsim}{ \mathop{}_{\textstyle \sim}^{\textstyle >} }
\newcommand{\lsim}{ \mathop{}_{\textstyle \sim}^{\textstyle <} }
\newcommand{\vev}[1]{ \left\langle {#1} \right\rangle }
\newcommand{\bra}[1]{ \langle {#1} | }
\newcommand{\ket}[1]{ | {#1} \rangle }
\newcommand{\EV}{ {\rm eV} }
\newcommand{\KEV}{ {\rm keV} }
\newcommand{\MEV}{ {\rm MeV} }
\newcommand{\GEV}{ {\rm GeV} }
\newcommand{\TEV}{ {\rm TeV} }
\def\diag{\mathop{\rm diag}\nolimits}
\def\Spin{\mathop{\rm Spin}}
\def\SO{\mathop{\rm SO}}
\def\O{\mathop{\rm O}}
\def\SU{\mathop{\rm SU}}
\def\U{\mathop{\rm U}}
\def\Sp{\mathop{\rm Sp}}
\def\SL{\mathop{\rm SL}}
\def\tr{\mathop{\rm tr}}

\def\IJMP{Int.~J.~Mod.~Phys. }
\def\MPL{Mod.~Phys.~Lett. }
\def\NP{Nucl.~Phys. }
\def\PL{Phys.~Lett. }
\def\PR{Phys.~Rev. }
\def\PRL{Phys.~Rev.~Lett. }
\def\PTP{Prog.~Theor.~Phys. }
\def\ZP{Z.~Phys. }

\newcommand{\ds}{\displaystyle}



\def\TODO#1{ {\bf ($\clubsuit$ #1 $\clubsuit$)} }


\baselineskip 0.7cm

\begin{titlepage}

\begin{flushright}
UT--11--48\\
IPMU--11--0215\\
\end{flushright}

\vskip 1.6cm
\begin{center}
{\large \bf
Higgs mass and muon anomalous magnetic moment\\ in the U(1) extended MSSM
}
\vskip 1.2cm

Motoi Endo$^{a,b}$,
Koichi Hamaguchi$^{a,b}$,
Sho Iwamoto$^{a}$,\\
Kazunori Nakayama$^{a,b}$
and
Norimi Yokozaki$^{a}$

\vskip 0.4cm

{\it $^a$Department of Physics, University of Tokyo, Tokyo 113-0033, Japan}\\
{\it $^b$Institute for the Physics and Mathematics of the Universe,
University of Tokyo, \\ Chiba 277-8583, Japan}\\

\vskip 1.5cm

\abstract{

We study phenomenological aspects of the MSSM with extra U(1) gauge symmetry.
We find that the lightest Higgs boson mass can be increased up to 125\,GeV, without introducing
a large SUSY scale or large $A$-terms, in the frameworks of the CMSSM and gauge mediated SUSY breaking (GMSB) models.
This scenario can simultaneously explain the discrepancy of the muon anomalous magnetic moment (muon $g-2$) at the $1\sigma$ level,
in both of the frameworks, U(1)-extended CMSSM/GMSB models.
In the CMSSM case, the dark matter abundance can also be explained.

}
\end{center}
\end{titlepage}

\setcounter{page}{2}


\section{Introduction}

Supersymmetry (SUSY) provides a natural solution to the hierarchy problem.
In the minimal SUSY standard model (MSSM),
the lightest Higgs boson mass is predicted to be lighter than the $Z$-boson at the tree level.
The radiative corrections make it heavier~\cite{Okada:1990vk},
and the LEP bound on the Higgs mass is avoided.

Recently, the ATLAS~\cite{ATLAS-CONF-2011-163}
and CMS~\cite{HIG-11-032} collaborations reported results of the searches for the standard model (SM)-like Higgs boson.
Both of them reported excesses of events,
which may be interpreted as signals of the SM-like Higgs boson whose mass is around 125\,GeV.
If the Higgs has such a mass indeed,
it provides critical information on the MSSM model,
since the Higgs boson mass significantly depends on the structure and parameters of the model.

A discrepancy of the experimental result~\cite{Bennett:2006fi} from the SM prediction of the anomalous magnetic moment ($g-2$) of the muon 
also indicates physics beyond the SM existing at TeV scale. The latest analyses of a hadronic contribution to the SM value 
provided the deviation at more than $3\sigma$ level \cite{g-2_hagiwara,g-2_davier}. This anomaly can be naturally explained in the 
SUSY models if the SUSY particles exist at around the 100\,GeV\,--\,1\,TeV scale.

SUSY predictions of the Higgs boson mass and the muon $g-2$ depend on soft SUSY breaking parameters, which are 
determined by the mediation mechanism of the SUSY breaking effect. In order to achieve the Higgs boson mass of 
$124\,\text{--}\,126$\,GeV, one needs a relatively large SUSY breaking mass scale and/or an appropriate size of the $A$-term of the 
top squark, whereas the soft mass scale is bounded from above to explain the muon $g-2$ anomaly. It is 
difficult to realize such a heavy Higgs boson with the muon $g-2$ result explained within the constrained MSSM (CMSSM)
and gauge mediated SUSY breaking (GMSB) models~\cite{Giudice:1998bp}, which are representative models of the SUSY breaking.

In this letter, we show that this frustration can be solved by 
an extension of the MSSM with an additional U(1) gauge symmetry.\footnote{MSSM with additional vector-like matters~\cite{Moroi:1991mg,vector_others,vector_g-2} can also explain the relatively heavy Higgs boson mass and the muon $g-2$ result simultaneously~\cite{vector_g-2} within the GMSB/CMSSM framework.}
The Higgs fields are charged under the symmetry, and the associated $D$-term provides an additional potential for the Higgs bosons.
It will be found that the Higgs boson mass can be as large as $124\,\text{--}\,126$\,GeV in a low soft mass scale even without a large $A$-term. In this parameter region the deviation of the muon $g-2$ can be explained by the SUSY contributions simultaneously.

There are many studies on the U(1) gauge extension of the SUSY models,
in particular, based on U(1)'s appearing in the grand unified theories 
(GUTs)~\cite{Langacker:2008yv,Han:2004yd}.
It was also pointed out that the additional $D$-term can raise the Higgs mass
even in the low-scale SUSY breaking models~\cite{Langacker:1999hs}.
Most of these studies are dedicated to solve the $\mu$-problem
and the matter content is rather complicated~\cite{Morrissey:2005uz}.
Here we consider simple U(1) extensions in order to make the discussion as clear and general as possible, 
which are sufficient for the purpose of enhancing the Higgs mass as well as explaining the muon $g-2$,
paying particular attention to the decoupling behavior of the $D$-term correction to the Higgs mass.
Although the similar topic was discussed in Ref.~\cite{Cincioglu:2010zz},
the decoupling effect was not properly taken into account.

In Sec.~\ref{sec:model} we describe our basic setup.
In Sec.~\ref{sec:ana} we 
perform a detailed analysis
in both the U(1)-extended CMSSM and GMSB models.
We explore the parameter regions where the lightest Higgs boson becomes as heavy as $124\,\text{--}\,126$\,GeV
and the observed muon $g-2$ is successfully explained.
We conclude this letter in Sec.~\ref{sec:conc}.
In Appendix, we show that the CMSSM models cannot explain the Higgs mass and the muon $g-2$ simultaneously 
even if we choose a large $A$-term, once the bound from $b\to s \gamma$ is imposed.
This may provide a motivation to introduce an extra U(1) gauge symmetry to raise the Higgs mass.

\section{The MSSM with extra U(1)}  \label{sec:model}

\subsection{Models of extra U(1)}

We consider an extension of the SM gauge groups to include additional U(1) gauge symmetry, U(1)$_X$.
There is one such anomaly-free U(1) known as U(1)$_{B-L}$, once the right-handed neutrinos are introduced.
In order to enhance the Higgs mass, however, the SM Higgs must have a charge of U(1)$_X$.
Thus U(1)$_{B-L}$ is not suitable for this purpose. Instead, U(1)$_X$ can be constructed as a linear combination of U(1)$_Y$ and U(1)$_{B-L}$. Such a gauge symmetry can be consistent with some GUT gauge groups, and 
various U(1) charge assignments are possible~\cite{Langacker:2008yv,Slansky:1981yr}.
In the minimal matter content, we consider two U(1) models, whose charge assignments are given in Table~\ref{table}.\footnote{
	If we extend the matter sector, many U(1) charge assignments can be obtained.
	A famous example is U(1)'s in the $E_6$ GUT, 
	which breaks, e.g., as $E_6 \to$ SO(10)$\times$ U(1) $\to$ SU(5)$\times$ U(1)$\times$ U(1).
	Although there are rich phenomenological implications in this kind of models, 
	we try to take the matter content as simple as possible.
	Also, there could be more complicated charge assignments if we allow 
	family non-universal U(1) symmetry~\cite{Chen:2008tc}.
}
The superpotential consists of
\begin{equation}
\begin{split}
	W_{\rm MSSM} &= y^{(d)}_{ij} Q_i\bar D_j H_d + y^{(u)}_{ij} Q_i\bar U_j H_u
	 + y^{(l)}_{ij} L_i\bar E_j H_d + y^{(\nu)}_{ij} L_i\bar N_j H_u
	+ \mu H_u H_d,\\
	W_S &= \lambda X(S\bar S-v^2),
	\label{super}
\end{split}
\end{equation}
and there are soft SUSY breaking terms.
The superpotential $W_S$ is introduced to break U(1)$_X$ spontaneously 
by vacuum expectation values (VEV's) of $S$ and $\bar S$, 
where $\lambda$ is a coupling constant, 
and $X$ is a singlet field under both the SM gauge groups and U(1)$_X$.

The first model in Table~\ref{table}, called U(1)$_\chi$, is motivated by the SO(10) GUT,
which has a breaking pattern like SO(10)$\to$ SU(5)$\times$ U(1)$_\chi$.
This extra U(1) is anomaly-free.
Here, the U(1)$_\chi$ charge assignments are taken to be 
consistent with the SO(10) embedding, 
and its gauge coupling constant $g_X$ is assumed to be unified with 
that of the SM gauge groups at the GUT scale, i.e., $g_X({\rm GUT})\simeq 0.7$, 
in the following numerical analysis. 
Hence the theory has $G_{\rm SM}\times$ U(1)$_\chi$ symmetry below the GUT scale,
where $G_{\rm SM}=$SU(3)$_c\times$ SU(2)$_L \times$ U(1)$_Y$ is the SM gauge groups, 
and U(1)$_\chi$ is assumed to be broken at around TeV scale by VEV's of $S$ and $\bar S$.

The next model is motivated by the Pati--Salam gauge group, 
SO(10)$\to$ SU(4)$\times$ SU(2)$_L\times$ SU(2)$_R$, 
where SU(2)$_R$ contains U(1) subgroup, generated by the $T_{3R}$ operator.
We regard this U(1)$_T$ as if it is the original symmetry of the theory,
and assume that it is finally broken by the VEV of $S$
without going into details of GUT constructions.
Similarly to the previous case,
this model has $G_{\rm SM}\times$ U(1)$_T$ symmetry below the GUT scale.
In the case of U(1)$_T$, we will not persist in constructing a full GUT theory.
In the following analysis, the U(1)$_T$ charges in Table~\ref{table} are taken 
to be twice as large as those in a GUT convention, and U(1)$_T$ gauge coupling constant 
is considered to be a free parameter rather than assumed to be unified with the SM gauge.
These two U(1)'s, U(1)$_\chi$ and U(1)$_T$, should be regarded as working examples of more broad classes of U(1) extensions.
Hereafter the new symmetry is represented by U(1)$_X$ whatever it is.

Some notes are in order. 
First of all, the right-handed neutrinos cannot have Majorana mass terms because of the U(1)$_X$ symmetry.
The seesaw mechanism may work at a TeV scale once the U(1)$_X$ symmetry is broken. 
For instance, a proper charge of $S$ could yield a Majorana mass term through the $S\bar N\bar N$ term 
after $S$ acquires a VEV if allowed by the U(1)$_X$ symmetry. 
Otherwise the neutrino mass purely comes from the Yukawa coupling.
Next, the $\mu$-term is allowed by the gauge symmetry. 
We implicitly assume some mechanism to solve the
$\mu$ problem. 
The $R$-symmetry or the Peccei--Quinn symmetry, or some discrete symmetry such as $Z_3$
may be used to forbid the $\mu$-term and to generate it dynamically.
Finally, it is assumed that $S$ and $\bar S$ are not in complete multiplets of SO(10),
and the parameter $y$ in their U(1)$_X$ charges is a free parameter.

\begin{table}[t]
  \begin{center}
    \begin{tabular}{  | c | c |  c  | c | c | }
      \hline 
         ~ & U(1)$_Y$ & U(1)$_{B-L}$ & 2$\sqrt{10}\times$U(1)$_\chi$  & U(1)$_T$ \\
       \hline 
       $Q$           & $1/6$  & $1/3$   & $-1$ & $0$ \\
       $\bar U$   & $-2/3$ & $-1/3$ & $-1$ & $-1$ \\ 
       $\bar D$   & $1/3$ & $-1/3$ & $3$ & $1$ \\ 
       $L$           & $-1/2$ & $-1$ & $3$ & $0$ \\
       $\bar E$   & $1$     & $1$   & $-1$ & $1$ \\ 
       $\bar N$   & $0$     & $1$   & $-5$ & $-1$ \\ \hline
       $H_u$      & $1/2$  & $0$   & $2$ & $1$ \\  
       $H_d$      & $-1/2$ & $0$   & $-2$ & $-1$ \\  \hline
       $S$           & $0$  & $0$   & $-y$ & $+y$ \\ 
       $\bar S$   & $0$  & $0$   & $+y$ & $-y$   \\ 
      \hline
    \end{tabular}
    \caption{ 
    		Anomaly-free U(1) charge assignments on the fields.
           }
    \label{table}
  \end{center}
\end{table}

\subsection{U(1)$_X$ contribution to Higgs mass and decoupling behavior}

When the Higgs fields are charged under U(1)$_X$, the associated $D$-term contributes to the Higgs quartic coupling. 
In the SUSY limit, 
this contribution decouples after the U(1)$_X$ gauge symmetry is broken. 
Thus, non-decoupling correction remains due to SUSY breaking effects~\cite{Batra:2003nj,Maloney:2004rc}.
This feature is taken into account by considering the whole U(1)$_X$ sector including the Higgs fields which 
break U(1)$_X$ spontaneously. 
The superpotential (\ref{super}) and the $D$-term of U(1)$_X$ as well as the SUSY breaking effect provide 
the scalar potential,
\begin{equation}
\begin{split}
	V_F &= |\lambda|^2| S\bar S - v^2 |^2 + |\lambda|^2 |X|^2 (|S|^2+|\bar S|^2),\\
	V_D &= \frac{1}{2}g_X^2\left[ x (|H_u|^2-|H_d|^2)+y(|S|^2-|\bar S|^2) \right]^2,\\
	V_{\rm SB} &= m_S^2 |S|^2 + m_{\bar S}^2 |\bar S|^2.
	\label{pot}
\end{split}
\end{equation}
Here $x$ denotes the U(1)$_X$ charge of $H_u$ and $H_d$, which is fixed to permit the Yukawa interactions of the matters (see Table \ref{table})
and $g_X$ is the gauge coupling constant of U(1)$_X$.

Let us find the minimum of the potential (\ref{pot}).
Under assumptions of $v \gg v_{H_u}, v_{H_d}$ and $m_S^2 = m_{\bar S}^2$, for simplicity,
the minimum is around $X=0$ and $v_S v_{\bar S} = (v^2-m_S^2/\lambda^2) \equiv \bar v^2$,
which are slightly shifted by $V_D$.
In the limit of $v_{H_u}=v_{H_d}=0$, a $D$-flat direction exists along $v_S = v_{\bar S}$,
whereas it is disturbed by finite $v_{H_u} = \langle H_u \rangle$ and $v_{H_d}= \langle H_d \rangle$.
Defining $v_S \equiv \bar v + \delta v_S$ and $v_{\bar S} \equiv \bar v+\delta v_{\bar S}$,
the true minimum is found as~\footnote{
	The SUSY breaking term also forces the minimum to be close to $v_S=v_{\bar S}$ as long as $m_S^2=m_{\bar S}^2$, 
	while $V_D$ tends to shift it towards $v_S \neq v_{\bar S}$ for $\tan \beta \neq 1$.
}
\begin{equation}
	\delta v_S \simeq -\delta v_{\bar S} \simeq -\frac{g_X^2 xy \bar v (|H_u|^2-|H_d|^2)}{2m_S^2 + m_{Z'}^2},
\end{equation}
where $m_{Z'}^2 = 4g_X^2 y^2 \bar v^2$ is a mass of the U(1)$_X$ boson.
Thus the scalar potential becomes~\footnote{
	One of the phase direction, $\arg(S) + \arg(\bar S)$, is fixed to be zero by minimizing $V_F$.
	The other combination is the Goldstone boson, which is eaten by the $Z'$ boson.
}
\begin{equation}
	V \simeq \frac{1}{2}g_X^2 x^2 \left( |H_u|^2-|H_d|^2\right)^2 \frac{2m_S^2}{2m_S^2 + m_{Z'}^2}.
\end{equation}
This serves an additional contribution to the Higgs potential arising at the tree level.
Then the following  terms are  added to the mass matrix of $(h_u^0, h_d^0)$;
\begin{equation}
	\Delta {\cal M}^2 \simeq g_X^2 x^2
	\begin{pmatrix}
		3v_{H_u}^2 - v_{H_d}^2 && -2v_{H_u} v_{H_d} \\
		-2v_{H_u} v_{H_d} && 	3v_{H_d}^2 - v_{H_u}^2
	\end{pmatrix}
	\frac{2m_S^2}{2m_S^2+m_{Z'}^2}.
\end{equation}
Consequently, the lightest Higgs boson mass receives the following correction 
\begin{equation}
	\Delta m_h^2 \simeq 2g_X^2x^2 (v_{H_u}^2+v_{H_d}^2)\cos^2(2\beta)\frac{2m_S^2}{2m_S^2+m_{Z'}^2}, \label{eq:dmh}
\end{equation}
in the limit $m_{A}^2 \gg m_Z^2$, where $m_A$ is the heavy CP-odd Higgs mass.

It is emphasized that the correction shows a decoupling behavior;
the correction disappears in the SUSY limit, i.e.~$m_S^2 / m_{Z'}^2 \to 0$~\cite{Batra:2003nj,Maloney:2004rc}.
In the CMSSM boundary condition, the soft mass, $m_S$, is correlated with the universal scalar mass $m_0$ or may be a free parameter, 
while in GMSB it is crucial that the messengers, $\Phi_{\rm mess}$ and $\bar \Phi_{\rm mess}$,
have the U(1)$_X$ charge, since otherwise $m_S$ is suppressed.
They will be discussed in Sec.~\ref{sec:CMSSM} and \ref{sec:GMSB}, respectively.

\subsection{U(1)$_X$ contribution to muon $g-2$}

The measurement of the muon $g-2$~\cite{Bennett:2006fi} shows a deviation from the SM prediction at more than the $3\sigma$ level as $\Delta a_\mu \equiv a_\mu({\rm exp}) - a_\mu({\rm SM}) = (26.1 \pm 8.0) \times 10^{-10}$~\cite{g-2_hagiwara,g-2_davier}. In the SUSY models, 
radiative corrections with superparticles can 
contribute to the magnetic moment. 
The SUSY contributions arise due to neutralino diagrams as well as those of the charginos.
Since the Higgs fields are charged both under the SM and U(1)$_X$ gauge symmetries, the neutralinos 
include the U(1)$_X$ gaugino and fermionic components of $S$, $\bar S$ and $X$. In the limit of $\lambda \gg g_X$, 
a couple of heavy components of the neutralinos are decoupled, which have a mass of order $\lambda \bar v$.
Then the mass matrix of the neutralinos becomes
\begin{equation}
{\cal M}_{\tilde\chi^0} =
\left(
\begin{array}{cccc|cc}
 & & & & 0 & 0 \\
 & & & & 0 & 0 \\
 & & {\rm MSSM} & & -\sqrt{2}g_X x v_d & 0 \\
 & & & & \sqrt{2}g_X x v_u & 0 \\ \hline
 0 & 0 & -\sqrt{2}g_X x v_d & \sqrt{2}g_X x v_u  & M_{\tilde Z'} & 2 g_X y \bar v\\
 0 & 0 & 0 & 0 & 2 g_X y \bar v & 0
\end{array}
\right)
\end{equation}
in a basis of $(\tilde B, \tilde W, \tilde H_d, \tilde H_u, \tilde Z', \tilde \Phi)$,
where $\tilde \Phi$ is the fermionic partner of the Goldstone boson which is absorbed into $Z'$, and $M_{\tilde Z'}$ is a 
SUSY breaking mass for the U(1)$_X$ gaugino.
The extra components of the neutralinos contribute to the muon $g-2$ through the mixing with the MSSM Higgsinos 
and couplings to the muon, since the left- and/or right-handed muons have a U(1)$_X$ charge. 

The U(1)$_X$ contributions are generally evaluated in the mass eigenstate basis (see e.g.,~\cite{Barger:2004mr}).
Noting that they mimic the Bino--smuon and Bino--Higgsino--smuon diagrams of the MSSM~\cite{Moroi:1995yh}, 
in the limit of $m_{Z'} \gg m_{\rm soft}, M_{\tilde Z'}$,
they are approximated as
\begin{eqnarray}
&& \Delta a_\mu^{{\rm U(1)}_X} \simeq 
-\frac{g_X^2}{8\pi^2} \frac{m_\mu^2 M_{\tilde Z'} \mu \tan\beta}{m_{Z'}^4} 
\left( Q_X^{L} Q_X^{\bar E} F_a(x) + Q_X^{L} Q_X^{H_u} F_b(x) + Q_X^{\bar E} Q_X^{H_u} F_b(x) \right),\\
&& F_a(x) = \frac{x^3+15x^2-9x-7-2(4x^2+7x+1)\ln x}{(x-1)^5},\\
&& F_b(x) = \frac{2(x^3+9x^2-9x-1-6x(x+1)\ln x)}{3(x-1)^5},
\end{eqnarray}
where $Q_X^i$ is a charge of the field, $i$, under U(1)$_X$ as provided in Table ~\ref{table}, and $x = m_{\rm soft}^2/m_{Z'}^2$ with a typical soft mass scale of the MSSM particles, $m_{\rm soft}$, including $\mu$.
Since $Z'$ must be heavier than $\sim$TeV from direct searches of $Z'$~\cite{Collaboration:2011dca,CMSPASEXO11019} 
and the electroweak precision bound~\cite{Cho:1998nr,Erler:2009jh}, 
the U(1)$_X$ contribution is found to be sufficiently suppressed 
even for a large $\tan\beta$. Thus, the SUSY prediction of the muon $g-2$ is determined by the MSSM contributions.

\section{Analysis}   \label{sec:ana}

\subsection{CMSSM}
\label{sec:CMSSM}

First, we analyze the Higgs mass and the muon $g-2$ in the CMSSM framework.
The boundary condition of the CMSSM framework is characterized by the five parameters, 
$(m_0, m_{1/2}, A_0, \tan \beta, {\rm sign}(\mu))$, while the soft scalar mass of $S (\bar S)$ in Eq.~(\ref{super}), $m_S ( = m_{\bar S})$, is chosen as a 
free parameter.
They except $\tan\beta$ are given at the GUT scale and then evolve following the renormalization group equations (RGEs)
toward the low energy.
We have adopted the SuSpect code~\cite{Djouadi:2002ze} for solving the RGEs,
which are modified to incorporate the effect of the additional U(1)$_X$ symmetry 
including a kinetic mixing between U(1)$_X$ and U(1)$_Y$ (see, e.g., Ref.~\cite{Babu:1996vt,Langacker:1998tc}) 
as well as the calculation of the mass spectrum of SUSY particles. The Higgs mass and the muon $g-2$ are calculated by FeynHiggs~\cite{Hahn:2010te}, and the relic abundance of the lightest neutralino is calculated by micrOMEGAs~\cite{Belanger:2010gh}. 
In the numerical analysis, uncertainties of the Higgs mass estimation are discarded unless otherwise mentioned, 
though the mass could shift by $\sim 2$\,GeV in the following figures. 

\begin{figure}[t]
\begin{center}
\includegraphics[scale=1.05]{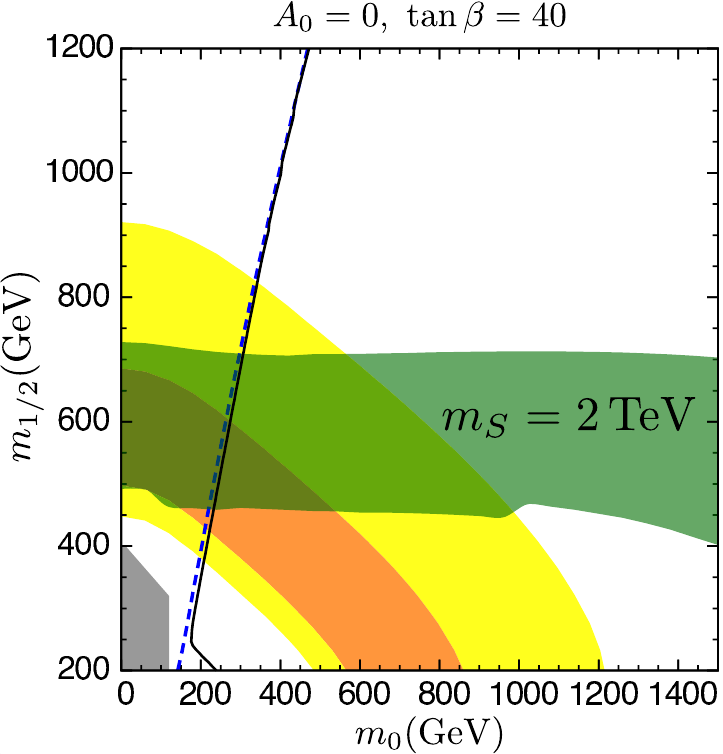}
\includegraphics[scale=1.05]{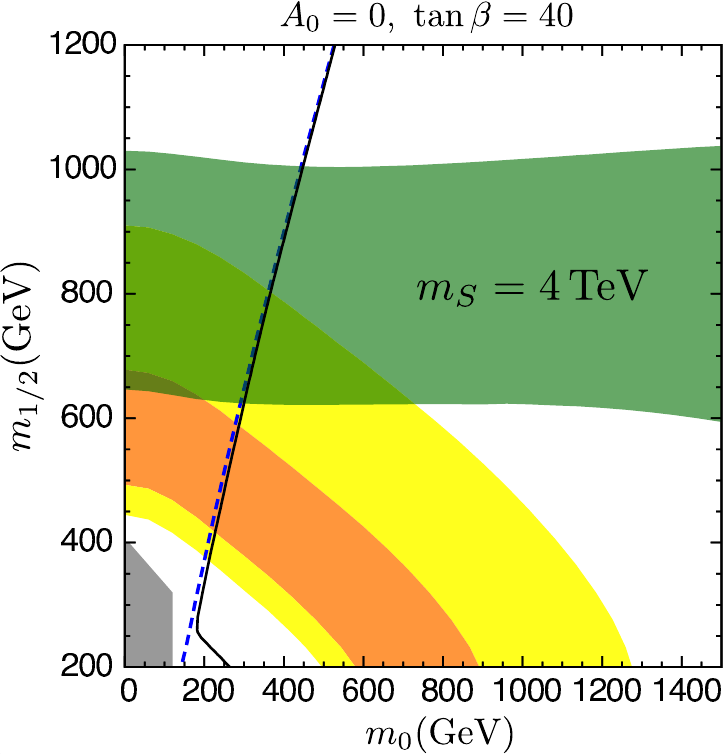}
\caption{Contours of the Higgs mass, the muon $g-2$ and the relic abundance of the dark matter
 in the CMSSM framework.
In the left (right) panel, U(1)$_T$ (U(1)$_\chi$) extension is considered. The region consistent with the Higgs mass, 124\,GeV $< m_h <$ 126\,GeV is shown as the green band for corresponding $m_S$. The orange (yellow) region is consistent with the muon $g-2$ at the 1$\sigma$ (2$\sigma$) level. The black solid line denotes the contour of the relic abundance of the lightest neutralino, $\Omega_{\rm CDM}h^2\simeq 0.11$.
The LSP is the lighter stau (the lightest neutralino) in the left (right) side of the blue dashed line.
The mass of the $Z'$ boson is set to be $m_{Z'}=2$\,TeV and CMSSM parameters are set to be $\tan \beta=40$, $A_0=0$ and ${\rm sign}(\mu)$=1 in both panels.
The U(1)$_X$ coupling constant is taken to coincide with SM gauge coupling constants at the GUT scale.
The gray region is excluded due to a tachyonic stau.
}
\label{fig:SUGRA}
\end{center}
\end{figure}

\begin{table}[t]
\begin{center}
\begin{tabular}{|c|c|c|c|}
\hline
 & CMSSM+$U(1)_T$ &  CMSSM+$U(1)_\chi$ & CMSSM
\\ \hline
$m_0$ & 300 & 300 &300\\
$m_{1/2}$ & 600  & 600 &600\\
 $A_0$ & 0 & 0 & 0\\
 $\tan\beta$ & 40 & 40 &40\\
 sign$(\mu)$ & +1 &  +1 & +1\\
 \hline \hline
 $\tilde{g}$ & 1391 & 1390 & 1395
 \\ \hline 
 $\tilde{q}_{1,L/R}$, $\tilde{q}_{2,L/R}$ & 1234--1287 & 1230--1289 &1254--1314
 \\
 $\tilde{q}_{3}$ & 971--1172 & 972--1175 & 990--1194
 \\ \hline
 $\tilde{\chi}^0_{3,4}$, $\tilde{\chi}^\pm_{2}$  & 700--716 & 689--706 &688--705
 \\
 $\tilde{\chi}^0_{2}$, $\tilde{\chi}^\pm_{1}$  & 506 & 505 & 506
 \\
 $\tilde{\chi}^0_1$ & 266 & 266 & 267
 \\ \hline
 $\tilde{e}_{L/R}$, $\tilde{\mu}_{L/R}$ & 402--511 & 386--524 & 463--575
 \\
 $\tilde{\tau}_1$ & 293 & 280 & 357
\\\hline
 $\tilde{N}$ & 337 & 395 & --
\\\hline
\end{tabular}
\caption{A comparison of the mass spectrum of the models.
}
\label{tab:mass}
\end{center}
\end{table}


The results are shown in Fig.~\ref{fig:SUGRA}, in the $(m_0, m_{1/2})$ planes,
for $\tan \beta=40$, $A_0=0$ and ${\rm sign}(\mu)$=1.
In this analysis the U(1)$_X$ coupling constant $g_X$ at the GUT scale is fixed to be the same as the SM gauge coupling constants.
In the left panel, the result for the U(1)$_T$ extension  with $m_S$=2\,TeV is shown.
The green band describes the parameter region
in which the Higgs boson mass is $124\,\text{--}\,126$\,GeV, and
the region consistent with the muon $g-2$ at the 1$\sigma$ (2$\sigma$) level
is shown by the orange (yellow) band.
Remarkably, the Higgs mass of $124\,\text{--}\,126$\,GeV and the muon $g-2$ (at the 1$\sigma$ level) can be 
simultaneously explained for $m_S \simeq 2$\,TeV in the U(1)$_T$ model when the charge assignment is provided
by Table~\ref{table}.

The black solid line denotes contours of the relic abundance of the lightest neutrino, 
$\Omega_{\rm CDM}h^2\simeq 0.11$, which is consistent with the WMAP observation~\cite{Komatsu:2010fb}. 
In the right region to the line, the abundance exceeds the measured dark matter abundance, whereas 
the coannihilation works in the region close to the blue dashed line, where the mass of the lightest neutralino 
equals to that of the lightest stau. 
It is emphasized that the dark matter abundance as well as the Higgs mass and the muon $g-2$ can be consistent with 
the experimental results, e.g.~for $m_0\simeq 300$\,GeV and $m_{1/2} \simeq 600$\,GeV.

A part of the relevant parameter region is already excluded by the LHC results~\cite{Aad:2011ib,CMS:SUS-11-008-pas}.
The exclusion can be inferred from the CMSSM results obtained by ATLAS and CMS, since 
the mass spectrum of the SUSY particles in the U(1) extended model is quite similar to that of the CMSSM.
Table ~\ref{tab:mass} shows the superparticle mass spectrum for $m_0\simeq 300$\,GeV and $m_{1/2} \simeq 600$\,GeV.
For comparison, 
the mass spectra for the cases of 
U(1)$_\chi$ and MSSM (no extra U(1)) are also shown.
It is seen that the mass spectrum is not much affected by the presence of the extra U(1)$_X$.
Applying the LHC exclusions to Fig.~\ref{fig:SUGRA}, in the region where the muon $g-2$ is consistent with the experimental value at the 1$\sigma$ level
and the LSP is the lightest neutralino
, the region with $m_{1/2}\lsim 500$\,GeV is already excluded~\cite{CMS:SUS-11-008-pas}.

We also show the result for the model with the $U(1)_{\chi}$ extension in the right panel of Fig.~\ref{fig:SUGRA}. 
The charge assignment and the gauge coupling constant of $U(1)_{\chi}$ are assumed to respect an underlying GUT.
It is found that the Higgs mass can be raised up to $124\,\text{--}\,126$\,GeV in the 2$\sigma$ region of the muon $g-2$ 
if the decoupling factor of the Higgs mass in (\ref{eq:dmh}) is almost maximized, e.g., as $2m_S^2/(2m_S^2 + m_{Z'}^2) \sim 0.9$ 
for $m_S = 4$\,TeV.

Let us compare the results with the CMSSM models which are discussed in Appendix.
If the Higgs fields are charged under the extra U(1) symmetry, the associated $D$-term can raise the Higgs mass 
without a large soft mass, so that the muon $g-2$ anomaly can be explained 
simultaneously. In particular, the trilinear couplings are set to vanish at the GUT scale, and thus,
the model is safe against the constraint from $b \to s\gamma$.

If $A_0$ including $A_t$ is enhanced to raise the top--stop contribution to the Higgs mass, the green regions in Fig.~\ref{fig:SUGRA} shift downwards.
On the other hand, the trilinear coupling of the stau tends to draw down the stau mass at the weak scale. Thus, the stau LSP region becomes wider, and 
it becomes difficult to explain the muon $g-2$ anomaly.
If $A_0$ becomes too large, $b \to s\gamma$ can be problematic similarly to the CMSSM.

In the figure, $\tan\beta$ was set to be 40. If it is increased, the Higgs mass decreases because of the bottom contribution to the Higgs mass, though the SUSY contributions to the muon $g-2$ are enhanced. On the other hand, when $\tan\beta$ is suppressed, the Higgs mass is lowered or stays unchanged so much, and
the muon $g-2$ becomes smaller. Thus, the current choice of $\tan\beta$ is almost 
the
best for the Higgs mass and the muon $g-2$.

If the $U(1)_X$ gauge coupling constant is larger than those of the SM gauge groups at the GUT scale, the extra contribution to the Higgs mass may be enhanced according to (\ref{eq:dmh}). 
However, this effect is small due to the RG evolution of the gauge coupling constants.

In the analysis, the mass of the $Z'$ boson was set to be $m_{Z'}=2$\,TeV. 
Note that this mass is large enough to satisfy the bounds from the direct searches of $Z'$~\cite{Collaboration:2011dca,CMSPASEXO11019} and the electroweak precision measurements~\cite{Cho:1998nr,Erler:2009jh}.
As the mass increases, the U(1)$_X$ $D$-term contribution to the Higgs potential becomes suppressed because of the decoupling behavior. We have checked that it is difficult to realize the Higgs mass of $124\,\text{--}\,126$\,GeV with the muon $g-2$ explained at the $2\sigma$ level for $m_{Z'} > 3$\,TeV as long as $m_S \lesssim 1$\,TeV and $A_0({\rm GUT}) = 0$.

\begin{figure}[t]
\begin{center}
\includegraphics[scale=0.96]{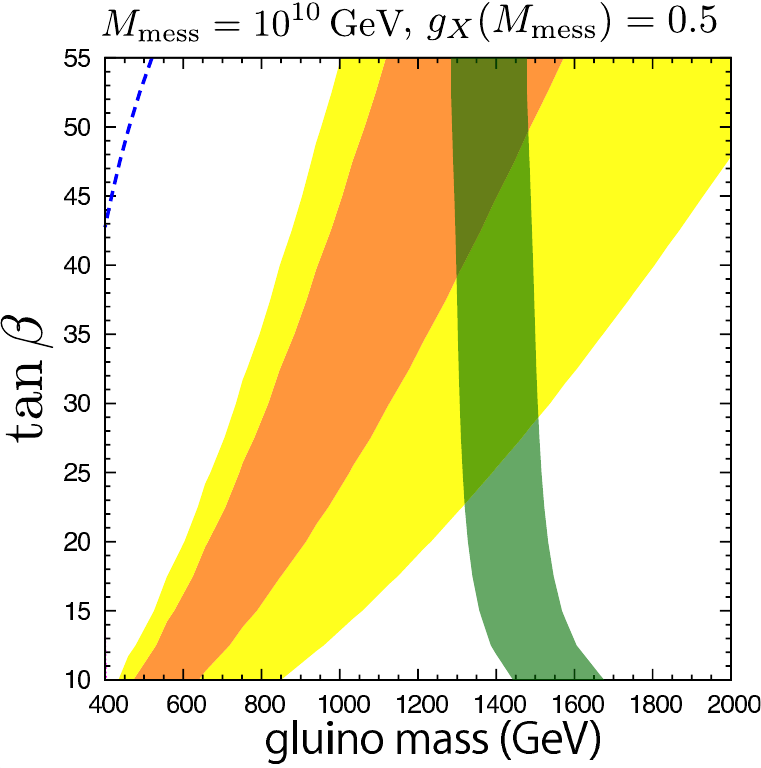}
\caption{
 	Contours of the Higgs mass and the muon $g-2$ in
        the U(1)$_T$-extended GMSB framework
	on the plane of the gluino mass and $\tan\beta$.
	The definition of each line is the same as that in Fig.~\ref{fig:SUGRA}.
	The mass of the $Z'$ boson is set to be $m_{Z'}=2$\,TeV, and the messenger scale is taken to be 
	$M_{\rm mess}=10^{10}$\,GeV.
	The U(1)$_X$ coupling constant is fixed to be $g_X(M_{\rm mess})=0.5$ at the messenger scale.
  }
\label{fig:GMSB}
\end{center}
\end{figure}

\subsection{GMSB}
\label{sec:GMSB}

Let us show the result for the case of GMSB. 
The messengers, $\Phi_{\rm mess}$ and $\bar\Phi_{\rm mess}$ are assumed to have U(1)$_X$ charges of $+n$ and $-n$
and ${\bf 5}$ and ${\bf \bar 5}$ representations under the SU(5), respectively.
For simplicity, $n=1$ is set in the following.
We introduce one such pair of $\Phi_{\rm mess}$ and $\bar\Phi_{\rm mess}$.
They couple to the SUSY breaking field $Z$ as
\begin{equation}
	W = Z \Phi_{\rm mess} \bar \Phi_{\rm mess}.
\end{equation}
The soft masses are obtained for $S$ and $\bar S$ through the U(1)$_X$ gauge interaction
at the messenger scale, $M_{\rm mess}$, as
\begin{equation}
	m_S^2 = m_{\bar S}^2 \simeq \left( \frac{g_X^2}{16\pi^2} \right)^2 10y^2 \Lambda^2,
\end{equation}
where $\Lambda \equiv F_Z/M_{\rm mess}$ is the soft SUSY breaking mass scale, which is around $100$\,TeV.
Note that all the matters receive similar corrections due to the U(1)$_X$ gauge interaction
depending on their U(1)$_X$ charges.

Results are shown in Fig.~\ref{fig:GMSB}.
In this analysis, $g_X$ is fixed to be $0.5$ at the messenger scale.
We show the contours of the Higgs mass and the muon $g-2$ in the U(1)$_T$ model
on the plane of the gluino mass and $\tan\beta$.
The definition of each line is the same as that in Fig.~\ref{fig:SUGRA}.
The mass of the $Z'$ boson is set to be $m_{Z'}=2$\,TeV, and the messenger scale is taken to be 
$M_{\rm mess}=10^{10}$\,GeV.
It is seen that the muon $g-2$ can be within the 1$\sigma$ range with $m_h = 124\,\text{--}\,126$\,GeV
for the gluino mass $\sim 1.4$\,TeV.
In the parameter region, the next-to-lightest SUSY particle is the neutralino.
Therefore, the model can be 
checked by searching for the SUSY event at the LHC accompanied by a large missing energy. 

Here, $g_X$ is set to be $0.5$ at the messenger scale. It is not unified with the SM gauge couplings at the GUT scale, 
and it is likely to blow up below the GUT scale since the messengers contribute to the gauge coupling evolutions above the messenger scale.\footnote{The blowing-up behavior may be ameliorated if the U(1) symmetry is embedded in 
a larger group
above the messenger scale.} If the coupling constant is assumed to be unified at the GUT scale, the U(1)$_X$ contribution to the Higgs mass is suppressed. To make matters worse, the messenger contribution to the soft mass of $S$ decreases. Consequently the decoupling behavior of the Higgs correction becomes more prominent. 
In particular,
the U(1)$_\chi$ setup, where the underlying GUT is respected especially for the gauge coupling constant, cannot enhance the Higgs mass large enough to explain the muon $g-2$ anomaly simultaneously. 

On the other hand, if $g_X$ is raised at the messenger scale discarding the blowing up, the Higgs can be heavier. However,
if it is too large,
 the electroweak symmetry breaking tends to be spoiled because the messenger contributes to the soft scalar mass of the up-type Higgs positively.

In the analysis, we chose a relatively high messenger scale. 
For a lower messenger scale,  the electroweak symmetry becomes unlikely to be broken, because 
the soft mass of
the up-type 
Higgs cannot evolve sufficiently during the RG running,
 and the up-type Higgs mass receives a positive contribution due to the extra U(1). Consequently, a high messenger scale is favored.

\section{Conclusion}   \label{sec:conc}

We have studied the U(1) gauge extensions of the MSSM
motivated by the recent results on the Higgs searches at the LHC,
which may indicate the Higgs boson whose mass is $\sim 125$\,GeV.
In the U(1) extended MSSM, the extra $D$-term gives an additional potential to the Higgs bosons
and hence the Higgs mass receives sizable corrections.
We have shown that this kind of models can explain the Higgs mass of around 125\,GeV
without introducing extremely heavy SUSY particles and/or a large $A$-term.
Furthermore, the anomaly of the muon $g-2$ can be explained 
at the 1$\sigma$ level simultaneously in the U(1)-extended CMSSM and GMSB.
The extra U(1) gauge boson mass is favored to be around a few TeV because of the 
decoupling behavior of the extra U(1) contribution to the Higgs mass. This mass region 
is expected to be covered by future LHC experiments~\cite{Aad:2009wy}.

\section*{Acknowledgment}

This work is supported by Grant-in-Aid for Scientific
research from the Ministry of Education, Science, Sports, and Culture
(MEXT), Japan, No.\ 21111006 (K.N.), and No.\ 22244030 (K.N.),
No. 23740172 (M.E.), 
No. 21740164 (K.H.), No. 22244021 (K.H.) and No. 22-7585 (N.Y.).
S.I. is supported by JSPS Grant-in-Aid for JSPS Fellows.
This work was supported by World Premier International Research Center Initiative (WPI Initiative), MEXT, Japan.

\begin{figure}[t]
\begin{center}
\includegraphics[scale=0.6]{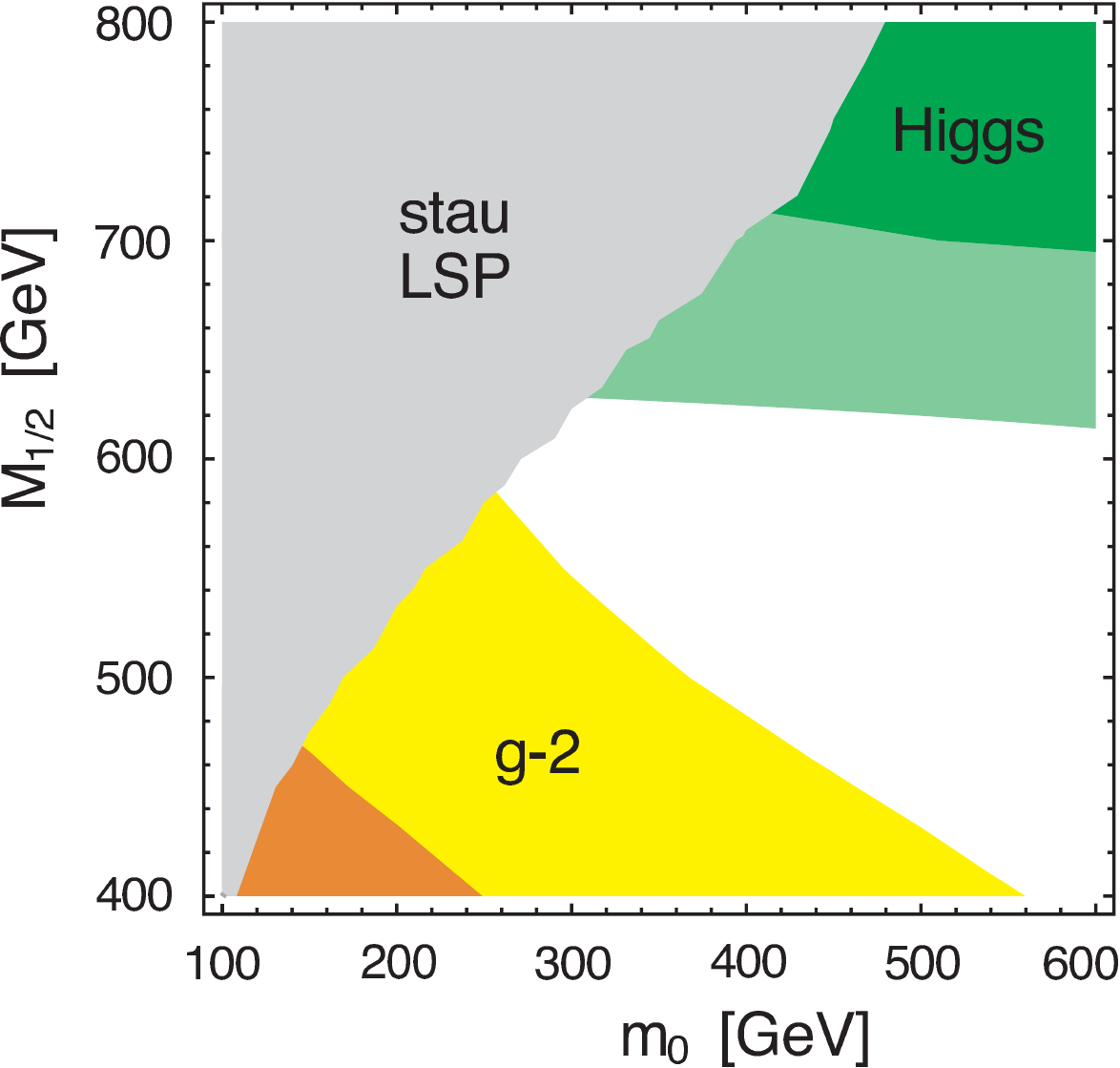}
\hspace*{5mm}
\includegraphics[scale=0.6]{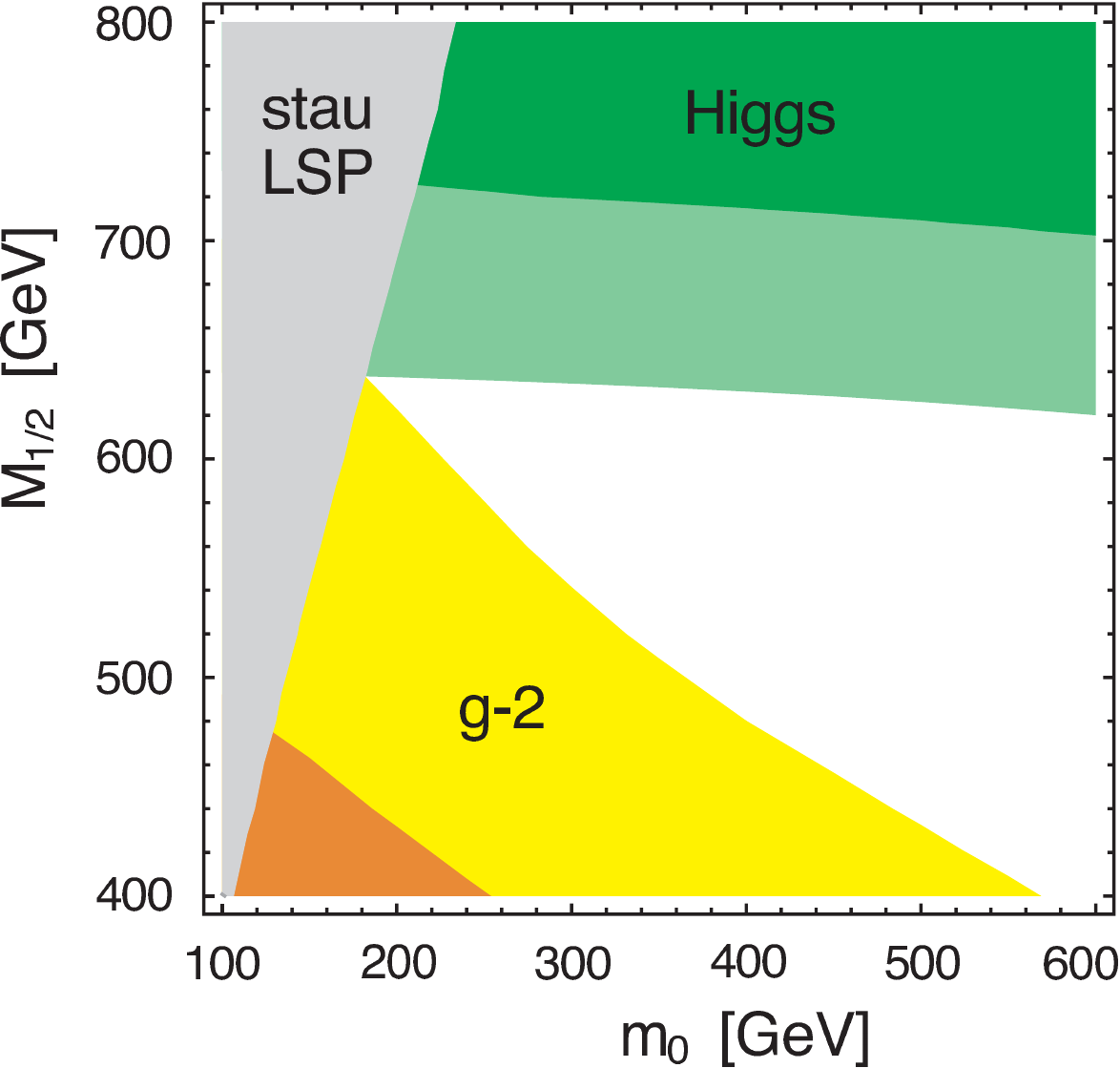}
\caption{%
Contours of the Higgs mass and the muon $g-2$ are shown. The Higgs mass are maximized by choosing $A_0$ and $A_u$ appropriately under the ${\rm Br}(\bar B \to X_s \gamma)$ constraint in the CMSSM models (left) and the extension (right), respectively (``$m_h$-max scenario''). In the dark green region, the Higgs mass is $124\,\text{--}\,126$\,GeV, and it becomes larger than 124\,GeV in the light green region once the uncertainties are included.
In the orange (yellow) regions, the muon $g-2$ is explained at the $1\sigma$ ($2\sigma$) level. The LSP is the (lighter) stau in the upper-left shaded region, while the lightest neutralino in the rest.
}
\label{fig:ATmax}
\end{center}
\end{figure}

\appendix
\section*{Appendix}

In this appendix we discuss the CMSSM models and their extension.
The CMSSM models have five input parameters, $(m_0, m_{1/2}, \tan\beta, {\rm sign}(\mu), A_0)$.
We consider an extended CMSSM framework where the trilinear couplings of the up-type squarks, $A_u$, are treated as a free parameter; it has six parameters, $(m_0, m_{1/2}, \tan\beta, {\rm sign}(\mu), A'_0, A_u)$, with $A'_0 \equiv A_e=A_d$.
The Higgs boson mass can be enhanced in a large trilinear coupling region of the top squark (``$m_h$-max scenario'').
However, we shall see that it is difficult to explain the Higgs mass of 125\,GeV and the muon $g-2$ anomaly simultaneously in the two framework.

One of the severest constraints comes from the branching ratio of the inclusive $\bar B \to X_s \gamma$ decay with $\bar B = \bar B^0$ or $B^-$. The experimental result, ${\rm Br}(\bar B \to X_s \gamma)^{\rm exp} = (3.55 \pm 0.24 \pm 0.09) \times 10^{-4}$~\cite{Asner:2010qj}, agrees well with the SM prediction, ${\rm Br}(\bar B \to X_s \gamma)^{\rm SM} = (3.15 \pm 0.23) \times 10^{-4}$~\cite{Misiak:2006zs}. Thus, the SUSY contribution is required to be in the range,
\begin{equation}
  -0.29 \times 10^{-4} < \Delta {\rm Br}(\bar B \to X_s \gamma) < 1.09 \times 10^{-4},
  \label{eq:bsg}
\end{equation}
at the $2\sigma$ level. Here, the errors are from the experimental and the SM uncertainties. In the analysis, the SUSY contributions are evaluated at the NLO level by SusyBSG~\cite{Degrassi:2007kj}. In addition to the uncertainties of the experimental value and the SM prediction in (\ref{eq:bsg}), extra errors of $10\%$ are taken into account both for the SUSY and charged Higgs contributions, respectively (see e.g.~\cite{Degrassi:2007kj}). It is found that the trilinear coupling of the top squark,
and thus
the Higgs boson mass, is bounded from above by ${\rm Br}(\bar B \to X_s \gamma)$.

In Fig.~\ref{fig:ATmax}, the Higgs mass and the muon $g-2$ are shown as contours in a $(m_0, m_{1/2})$ plane with $(\tan\beta, {\rm sign}(\mu)) = (20, +1)$.
The renormalization group equations are solved and the mass spectrum of the superparticles are evaluated by SoftSUSY~\cite{Allanach:2001kg}.
The Higgs mass is obtained by using FeynHiggs~\cite{Hahn:2010te}.
Uncertainties of the Higgs mass estimation is also taken into account with relying on FeynHiggs.
The left panel is the result for the CMSSM framework, and the value of $A_0$ $(=A_u=A_d=A_e)$ is tuned so that the Higgs mass is maximized under the constraint of ${\rm Br}(\bar B \to X_s \gamma)$.
In the right panel $A'_0$ $(=A_d=A_e)$ is set to be zero and $A_u$ is appropriately tuned as is done in the left panel.

In the dark green regions, the Higgs mass is calculated as large as $m_h=124\,\text{--}\,126$\,GeV.
In the light green region, the Higgs mass can be larger than 124\,GeV if the theoretical uncertainties are included. On the other hand, the muon $g-2$ estimated by FeynHiggs is explained within the $1\sigma$ ($2\sigma$) levels in the orange (yellow) region. The upper-left gray region is forbidden because of the stau LSP, while just below it is the coannihilation region.

It is seen from the left panel of Fig.~\ref{fig:ATmax} that a large part of the small $m_0$ region is excluded by the LSP stau in the CMSSM setup. If the universality of the trilinear coupling is violated  as in the right panel, the region of the Higgs mass of $124\,\text{--}\,126$\,GeV can approach to that favored by the muon $g-2$ significantly. 
Nonetheless, the Higgs mass of 124\,GeV and an explanation of the muon $g-2$ anomaly (at the $\sim 2\sigma$ level) cannot be simultaneously achieved. 

The situation is not improved for different choices of $\tan\beta$. If it is increased, the bound from ${\rm Br}(\bar B \to X_s \gamma)$ becomes severer, and the muon $g-2$ decreases for smaller $\tan\beta$. In both cases, the separation between the regions favored by the Higgs mass and the muon $g-2$ turns out to be wider. 

The main reason for the difficulty of the above result is that the constraint from $\bar B \to X_s \gamma$ sets an upper bound
on the parameter $A_t$, and consequently the Higgs mass is bounded from above. If the soft scalar mass of the up- and down-type Higgses are assumed to be non-universal against $m_0$, the $\bar B \to X_s \gamma$ bound can be relaxed, while attention should be paid for other constraints such as ${\rm Br}(B_s \to \mu\mu)$.



\begin{thebibliography}{99}


\bibitem{Okada:1990vk}
Y.~Okada, M.~Yamaguchi and T.~Yanagida,
Prog.\ Theor.\ Phys.\ {\bf 85} (1991) 1;
Phys.\ Lett.\ B {\bf 262} (1991) 54;
J.~R.~Ellis, G.~Ridolfi and F.~Zwirner,
Phys.\ Lett.\ B {\bf 257} (1991) 83;
H.~E.~Haber and R.~Hempfling,
Phys.\ Rev.\ Lett.\ {\bf 66} (1991) 1815.

\bibitem{ATLAS-CONF-2011-163}
  ATLAS NOTE, ATLAS-CONF-2011-163.

\bibitem{HIG-11-032}
  CMS Physics Analysis Summary, HIG-11-032.

\bibitem{Bennett:2006fi} 
  G.~W.~Bennett {\it et al.}  [Muon G-2 Collaboration],
  Phys.\ Rev.\ D {\bf 73}, 072003 (2006)
  [hep-ex/0602035].

\bibitem{g-2_hagiwara}
  K.~Hagiwara, A.~D.~Martin, D.~Nomura, T.~Teubner,
  Phys.\ Lett.\  {\bf B649}, 173-179 (2007).
  [hep-ph/0611102];
  T.~Teubner, K.~Hagiwara, R.~Liao, A.~D.~Martin, D.~Nomura,
    [arXiv:1001.5401 [hep-ph]];
  K.~Hagiwara, R.~Liao, A.~D.~Martin, D.~Nomura, T.~Teubner,
  J.\ Phys.\ G {\bf G38}, 085003 (2011).
  [arXiv:1105.3149 [hep-ph]].


\bibitem{g-2_davier}
  M.~Davier, A.~Hoecker, G.~Lopez Castro, B.~Malaescu, X.~H.~Mo, G.~Toledo Sanchez, P.~Wang, C.~Z.~Yuan {\it et al.},
  Eur.\ Phys.\ J.\  {\bf C66}, 127-136 (2010).
  [arXiv:0906.5443 [hep-ph]];
  M.~Davier, A.~Hoecker, B.~Malaescu, C.~Z.~Yuan, Z.~Zhang,
  Eur.\ Phys.\ J.\  {\bf C66}, 1-9 (2010).
  [arXiv:0908.4300 [hep-ph]];
  M.~Davier, A.~Hoecker, B.~Malaescu, Z.~Zhang,
  Eur.\ Phys.\ J.\  {\bf C71}, 1515 (2011).
  [arXiv:1010.4180 [hep-ph]].


\bibitem{Giudice:1998bp} 
  G.~F.~Giudice and R.~Rattazzi,
  Phys.\ Rept.\  {\bf 322}, 419 (1999)
  [hep-ph/9801271].
  
  
\bibitem{Moroi:1991mg}
  T.~Moroi and Y.~Okada,
  Mod.\ Phys.\ Lett.\ A {\bf 7} (1992) 187;
  Phys.\ Lett.\ B {\bf 295} (1992) 73.

  \bibitem{vector_others}
  M.~Asano, T.~Moroi, R.~Sato and T.~T.~Yanagida,
  Phys.\ Lett.\ B {\bf 705}, 337 (2011)
  [arXiv:1108.2402 [hep-ph]];
  J.~L.~Evans, M.~Ibe and T.~T.~Yanagida,
  arXiv:1108.3437 [hep-ph].


\bibitem{vector_g-2}
  M.~Endo, K.~Hamaguchi, S.~Iwamoto and N.~Yokozaki,
  Phys.\ Rev.\ D {\bf 84}, 075017 (2011)
  [arXiv:1108.3071 [hep-ph]];
  arXiv:1112.5653 [hep-ph];
  T.~Moroi, R.~Sato and T.~T.~Yanagida,
  arXiv:1112.3142 [hep-ph].

  
    
\bibitem{Langacker:2008yv}
  See for a review, P.~Langacker,
  Rev.\ Mod.\ Phys.\  {\bf 81}, 1199 (2009)
  [arXiv:0801.1345 [hep-ph]].

\bibitem{Han:2004yd}
  T.~Han, P.~Langacker and B.~McElrath,
  Phys.\ Rev.\  D {\bf 70}, 115006 (2004)
  [arXiv:hep-ph/0405244];
  S.~F.~King, S.~Moretti and R.~Nevzorov,
  Phys.\ Rev.\  D {\bf 73}, 035009 (2006)
  [arXiv:hep-ph/0510419];
  Phys.\ Lett.\  B {\bf 634}, 278 (2006)
  [arXiv:hep-ph/0511256];
  V.~Barger, P.~Langacker, H.~S.~Lee and G.~Shaughnessy,
  Phys.\ Rev.\  D {\bf 73}, 115010 (2006)
  [arXiv:hep-ph/0603247];
  R.~Howl and S.~F.~King,
  JHEP {\bf 0801}, 030 (2008)
  [arXiv:0708.1451 [hep-ph]];
  T.~Cohen and A.~Pierce,
  Phys.\ Rev.\  D {\bf 78}, 055012 (2008)
  [arXiv:0803.0765 [hep-ph]];
  H.~Sert, E.~Cincioglu, D.~A.~Demir and L.~Solmaz,
  Phys.\ Lett.\ B {\bf 692}, 327 (2010)
  [arXiv:1005.1674 [hep-ph]].
  
\bibitem{Langacker:1999hs} 
  P.~Langacker, N.~Polonsky and J.~Wang,
  Phys.\ Rev.\ D {\bf 60}, 115005 (1999)
  [hep-ph/9905252].

    
\bibitem{Morrissey:2005uz} 
  D.~E.~Morrissey and J.~D.~Wells,
  Phys.\ Rev.\ D {\bf 74}, 015008 (2006)
  [hep-ph/0512019].
  
\bibitem{Cincioglu:2010zz} 
  E.~Cincioglu, Z.~Kirca, H.~Sert, S.~Solmaz, L.~Solmaz and Y.~Hicyilmaz,
  Phys.\ Rev.\ D {\bf 82}, 055009 (2010).
  
\bibitem{Slansky:1981yr}
  R.~Slansky,
  Phys.\ Rept.\  {\bf 79}, 1 (1981).
  
\bibitem{Chen:2008tc}
  M.~C.~Chen, D.~R.~T.~Jones, A.~Rajaraman and H.~B.~Yu,
  Phys.\ Rev.\  D {\bf 78}, 015019 (2008)
  [arXiv:0801.0248 [hep-ph]].
  
\bibitem{Batra:2003nj} 
  P.~Batra, A.~Delgado, D.~E.~Kaplan and T.~M.~P.~Tait,
  JHEP {\bf 0402}, 043 (2004)
  [hep-ph/0309149].
  
\bibitem{Maloney:2004rc} 
  A.~Maloney, A.~Pierce and J.~G.~Wacker,
  JHEP {\bf 0606}, 034 (2006)
  [hep-ph/0409127].

\bibitem{Barger:2004mr} 
  V.~Barger, C.~Kao, P.~Langacker and H.~-S.~Lee,
  Phys.\ Lett.\ B {\bf 614}, 67 (2005)
  [hep-ph/0412136].

\bibitem{Moroi:1995yh} 
  T.~Moroi,
  Phys.\ Rev.\ D {\bf 53}, 6565 (1996)
  [Erratum-ibid.\ D {\bf 56}, 4424 (1997)]
  [hep-ph/9512396].
  
\bibitem{Collaboration:2011dca} 
  {\bf ATLAS} Collaboration, 
  arXiv:1108.1582 [hep-ex].

\bibitem{CMSPASEXO11019}
{\bf CMS} Collaboration, CMS PAS EXO-11-019.
  
\bibitem{Cho:1998nr} 
  G.~-C.~Cho, K.~Hagiwara and Y.~Umeda,
  Nucl.\ Phys.\ B {\bf 531}, 65 (1998)
  [Erratum-ibid.\ B {\bf 555}, 651 (1999)]
  [hep-ph/9805448].
  
\bibitem{Erler:2009jh} 
  J.~Erler, P.~Langacker, S.~Munir and E.~Rojas,
  JHEP {\bf 0908}, 017 (2009)
  [arXiv:0906.2435 [hep-ph]].
  
\bibitem{Djouadi:2002ze} 
  A.~Djouadi, J.~-L.~Kneur and G.~Moultaka,
  Comput.\ Phys.\ Commun.\  {\bf 176}, 426 (2007)
  [hep-ph/0211331].
  
\bibitem{Babu:1996vt} 
  K.~S.~Babu, C.~F.~Kolda and J.~March-Russell,
  Phys.\ Rev.\ D {\bf 54}, 4635 (1996)
  [hep-ph/9603212].
  
\bibitem{Langacker:1998tc} 
  P.~Langacker and J.~Wang,
  Phys.\ Rev.\ D {\bf 58}, 115010 (1998)
  [hep-ph/9804428].
  
\bibitem{Hahn:2010te}
  T.~Hahn, S.~Heinemeyer, W.~Hollik, H.~Rzehak and G.~Weiglein,
  Nucl.\ Phys.\ Proc.\ Suppl.\  {\bf 205-206}, 152 (2010)
  [arXiv:1007.0956 [hep-ph]].
  
\bibitem{Belanger:2010gh}
  G.~Belanger, F.~Boudjema, P.~Brun, A.~Pukhov, S.~Rosier-Lees, P.~Salati and A.~Semenov,
  Comput.\ Phys.\ Commun.\  {\bf 182}, 842 (2011)
  [arXiv:1004.1092 [hep-ph]].
  
\bibitem{Komatsu:2010fb} 
  E.~Komatsu {\it et al.}  [WMAP Collaboration],
  Astrophys.\ J.\ Suppl.\  {\bf 192}, 18 (2011)
  [arXiv:1001.4538 [astro-ph.CO]].
  
\bibitem{Aad:2011ib} 
  G.~Aad {\it et al.}  [ATLAS Collaboration],
  arXiv:1109.6572 [hep-ex].
  
\bibitem{CMS:SUS-11-008-pas}
  The CMS Collaboration,
  CMS PAS SUS--11--008.

\bibitem{Aad:2009wy}
  G.~Aad {\it et al.}  [The ATLAS Collaboration],
  arXiv:0901.0512 [hep-ex].
  
\bibitem{Asner:2010qj} 
  D.~Asner {\it et al.}  [Heavy Flavor Averaging Group Collaboration],
  arXiv:1010.1589 [hep-ex].

\bibitem{Misiak:2006zs} 
  M.~Misiak, H.~M.~Asatrian, K.~Bieri, M.~Czakon, A.~Czarnecki, T.~Ewerth, A.~Ferroglia and P.~Gambino {\it et al.},
  Phys.\ Rev.\ Lett.\  {\bf 98}, 022002 (2007)
  [hep-ph/0609232].

\bibitem{Degrassi:2007kj} 
  G.~Degrassi, P.~Gambino and P.~Slavich,
  Comput.\ Phys.\ Commun.\  {\bf 179}, 759 (2008)
  [arXiv:0712.3265 [hep-ph]].

\bibitem{Allanach:2001kg}
  B.~C.~Allanach,
  Comput.\ Phys.\ Commun.\  {\bf 143}, 305 (2002)
  [arXiv:hep-ph/0104145].

\end{thebibliography}
\end{document}